\documentclass{aa}
\usepackage{psfig}

\newcommand{\delay}	{-0.11}		
\newcommand{\errdelay}	{0.01}		
\newcommand{\rc} 	{37.8}		
\newcommand{\errrc}	{0.6} 		
\newcommand{\conc}	{1.08} 		
\newcommand{\errc}	{0.02}		
\newcommand{\rcup}	{+3.30}		
\newcommand{\rclow}	{-5.16} 	
\newcommand{\concup}	{+0.33}		
\newcommand{\conclow}	{-0.08}		
\newcommand{\dx}	{3} 		
\newcommand{\dy}	{-5} 		
\newcommand{\mvhba} 	{0.78}		
\newcommand{\errmva}	{0.05}		
\newcommand{\mvhbb} 	{0.67}		
\newcommand{\errmvb}	{0.05}		
\newcommand{\mmzero} 	{16.45} 	
\newcommand{\errmm} 	{0.10} 		
\newcommand{\vhb} 	{17.18} 	
\newcommand{\errvhb} 	{0.02} 		
\newcommand{\apparent} 	{16.51} 	
\newcommand{\ds}	{19.5}		
\newcommand{\eds}	{0.9}		
\newcommand{\dgc}	{16.2}		
\newcommand{\edgc}	{0.7}		
\newcommand{\dgp}	{14.4}		
\newcommand{\edgp}	{0.7}		

\begin{document}
\thesaurus{	08.08.1; 	 	
		08.16.3; 	 	
		10.07.3 Palomar 12 	
	 }
\title	{	Young Galactic Globular Clusters II. The case of Pal~12.
\thanks	{	Based on data collected at the European Southern Observatory, 
		La Silla, Chile}
	}
\author {	A. Rosenberg 	\inst{1} \and 
		I. Saviane 	\inst{2} \and 
		G. Piotto 	\inst{2} \and 
		E.V. Held 	\inst{3}
	}
\offprints {	A. Rosenberg: alf@obelix.pd.astro.it	}
\institute{	
		Telescopio Nazionale Galileo, 
		vicolo dell'Osservatorio 5, I--35122 Padova, Italy
\and
		Dipartimento di Astronomia, Univ. di Padova, vicolo
		dell'Osservatorio 5, I--35122 Padova, Italy
\and
		Osservatorio Astronomico di Padova, 
		vicolo dell'Osservatorio 5, I--35122 Padova, Italy
	}
\date{}
\titlerunning {Young GGCs (II): The case of Palomar 12}
\maketitle

\begin{abstract}

We present broadband {\it V,I} CCD photometry for $\sim1700$ stars
towards the Galactic globular cluster Palomar~12, covering a field of
$10\farcm 7 \times 10\farcm 7$. From these data, a color-magnitude
diagram from the red giant branch tip to $\sim2$ mag below the
cluster's turn-off is obtained. From a comparison with the color
magnitude diagrams of 47~Tuc and M5, and using different theoretical
models, we confirm that Pal~12 is younger, finding an age $68\pm10\%$
that of both template clusters. Revised structural parameters are also
obtained: $r_{\rm c} = \rc \pm \errrc$ and $c = \conc \pm \errc$.

\end{abstract}

\section{Introduction} \label{intro}

The formation of the Galactic halo is presently at the center of an
open debate. Stetson et al. (\cite{s_96}) state that, apart from a
handful of anomalous clusters that may well have been captured from a
satellite dwarf galaxy, there is no strong evidence for a significant
spread in age among clusters of a given metal abundance, while
Chaboyer et al. (\cite{c_96}) support an age spread of 5 Gyr among the
bulk of the Galactic globular clusters (GGCs) (which is increased to 9
Gyr, if the youngest clusters are considered).

One of the largest underlying sources of uncertainty is the
heterogeneity of the data used in these studies, which prevents
``large scale'' tests. This is the main reason that prompted our group
to gather an homogeneous photometric data base of GGC, in $V$ and $I$,
as discussed in Saviane et al. (\cite{s97}). To date, we have observed
about 80\% of the closest GGC's ($(m-M)_v<16$) with 1m class
telescopes, and our data set allows us to obtain color-magnitude
diagrams (CMDs) from the RGB tip down to a few magnitudes below the
turn-off (TO).

Several young (or suspect young) GGCs were included in our program:
Palomar~12 (Gratton \& Ortolani \cite{g88}, hereafter GO88),
Ruprecht~106 (Buonanno et al.\cite{b90}), Arp~2 (Buonanno et
al. \cite{b95a}), Terzan~7 (Buonanno et al. \cite{b95b}) and Palomar~1
(Rosenberg et al. \cite{r98}, hereafter Paper I). Their age is
suspected to significantly deviate from the general distribution. A
precise determination of this deviation within our homogeneous data
set is particularly valuable in the general framework of the GGC ages
and of great importance in order to decide on the models of Galactic
formation. For this reason, we will present and discuss in separate
papers of this series the photometric data for the clusters whose age
is significantly different from the average age of the GGC in our
sample.

In Paper I we have discussed the case of Palomar 1, which resulted to
be the youngest GGC in our Galaxy. In this paper we concentrate on
Pal~12 (C2143-214, $\alpha_{\rm 2000}=21^h\,46\fm6$, $\delta_{\rm
2000}=-21^\circ\,15'$; l$=30\fdg5$, b$=-47\fdg7$) discovered by
Harrington \& Zwicky (\cite{h53}) on the Palomar Sky Survey plates.
Indeed, Pal~12 has been the first cluster to be classified as younger
than the bulk of GGCs. However, the age determination in previous
studies of its CMD, namely Stetson et al. (\cite{s89}, hereafter S89),
and Da Costa \& Armandroff (\cite{d90}, hereafter DA90) was affected
by large uncertainties in its metal content (more than 0.3 dex in
[Fe/H]). Since then, new low- and high-resolution spectroscopy have
been used in order to estimate the metallicity of Palomar~12
(Armandroff \& Da Costa \cite{a91}, AD91; Brown et al. \cite{b97},
B97).

The observations and data reduction are presented in
Sect.~\ref{obs_reduct} and the resulting CMD is discussed in
Sect.~\ref{sec_cmd}. The relative age determination is carried out in
Sect.~\ref{age}, and in Sect.~\ref{sec_struct} the structural
parameters of the cluster are established.

\section{Observations and data reductions}
\label{obs_reduct}

\begin{figure*} 
\psfig{figure=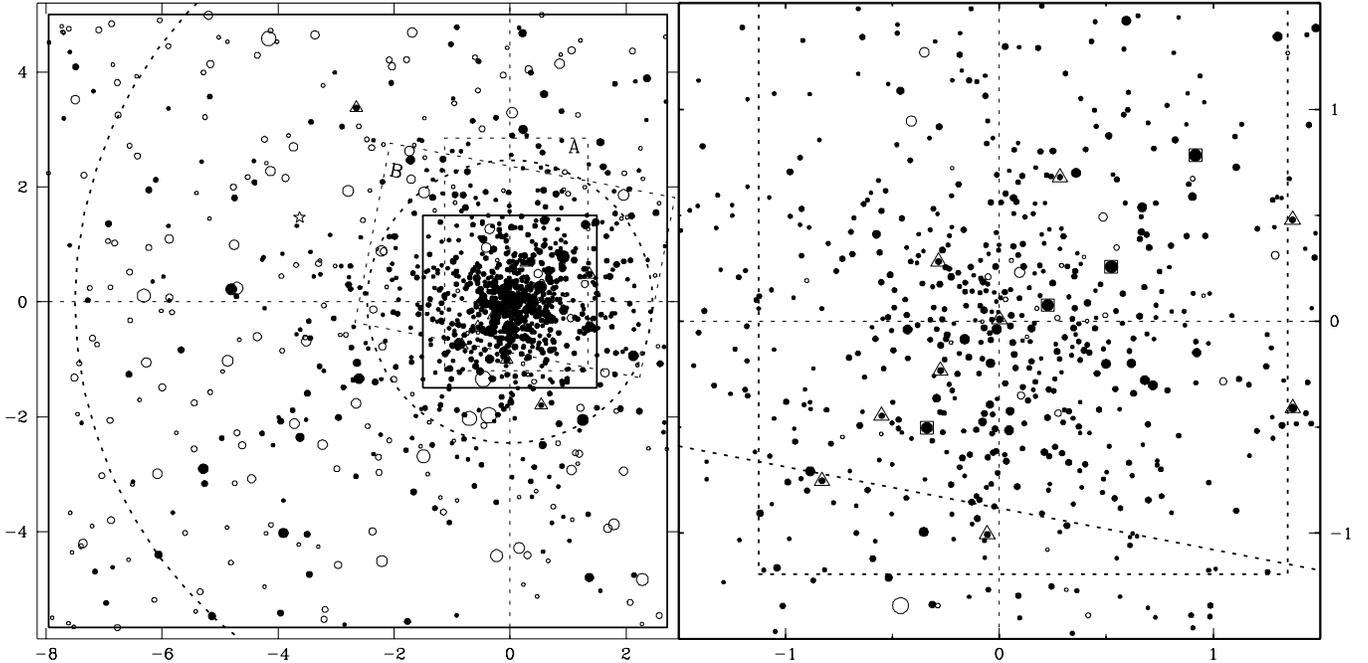,width=18cm}
\caption[]{Schematic representation of the complete ($10\farcm7\times
10\farcm7$, {\it left}), and the central ($1\farcm5\times1\farcm5$,
{\it right}) field, where stars brighter than $V\sim22.5$ are plotted.
Coordinates are in arcmin (North is at the top, West to the right).
Filled circles represent the stars within $3\sigma$ in color from the
fiducial sequence of the CMD plus BSS, HB and blended stars (see text
for a detailed explanation). BSS are also marked as triangles, while
squares indicate the stars for which spectroscopy has been
obtained. Also, fields covered by GO88 (A) and S89 (B) are indicated
(dashed rectangles). The tidal radius obtained by Trager et al, 1995
($2\farcm 46$) is represented together with our new estimate based on
the present data. A spiral galaxy is also clearly identified in our
field, marked with a five-pointed star}
\label{field1}
\end{figure*}

The data were collected on October 20, 1997 at the ESO Danish~1.54m
telescope equipped with DFOSC. The camera employed a $2048\times2048$
pixels Loral CCD, with a pixel size of 0\farcs40 on the sky, for a
total field of view of $13\farcm6\times 13\farcm 6$.

Tab.~\ref{obs} lists the complete log of the observations. The
weather conditions were good during the night, which was stable and
photometric, and the seeing was $\sim1\arcsec$.

\begin{table}
\caption[]{Journal of the Pal~12 observations for Oct 20, 1995}
\label{obs}
\begin{tabular}{ccccc}
\hline
\noalign{\smallskip}
 ID & Filter & $t_{\rm exp}$ (s) & $X$ & FWHM\arcsec \\
\noalign{\smallskip}
\hline 
\noalign{\smallskip}
 V1 & V & 20 & 1.010 & 0.9 \\ 
 V2 & V & 20 & 1.012 & 1.0 \\ 
 V3 & V & 40 & 1.013 & 1.0 \\ 
 V4 & V & 50 & 1.015 & 1.0 \\ 
 V5 & V & 50 & 1.016 & 1.0 \\ 
 V6 & V & 300 & 1.032 & 1.1 \\ 
 V7 & V & 300 & 1.040 & 1.1 \\ 
 I1 & I & 30 & 1.091 & 1.0 \\ 
 I2 & I & 40 & 1.086 & 1.0 \\ 
 I3 & I & 40 & 1.082 & 0.9 \\ 
 I4 & I & 40 & 1.074 & 0.9 \\ 
 I5 & I & 300 & 1.057 & 1.0 \\ 
 I6 & I & 300 & 1.049 & 1.0 \\ 
\noalign{\smallskip}
\hline
\end{tabular}
\end{table}

The image processing was carried out within the {\sc iraf}
environment. First, a map of the bad features of the chip was created
and they were removed from the raw images. Then, the bias stability
was checked by comparing frames taken at different times during the
entire run, and no significant discrepancies were found. A 0.4~\%
spatial gradient was found along the $x$ direction, thus a master bias
image was created by taking the median of all the bias images. This
master bias image was subtracted from all the remaining frames.

Sky flats were used to create master flat fields as medians of the
single frames.

In order to avoid the fall of quantum efficiency (QE) all around the
border of the Loral CCD, we cut our images outside the limit where the
QE was 90\% of the central value. From an inspection of the flats this
limit imposed an effective area of $1600\times 1600$~pixels
(i.e. $10\farcm 7 \times 10\farcm 7$; Saviane \& Held \cite{s98},
hereafter SH98, give further details). The effective field is
schematically represented in Fig.~\ref{field1}.

Stellar photometry was performed using {\sc daophot}, {\sc allstar}
(Stetson, \cite{s87}), and {\sc allframe}, according to a standard
procedure (see Paper I).

Observations of Landolt's (\cite{l92}) standard stars were used to
calibrate the photometry. In addition, the shutter delay time was
measured with a sequence of images taken with increasing exposure
times. A value of $\delay~\pm\errdelay$s (where the error is the
standard deviation) was found. The raw magnitudes were first
normalized according to the following equation
\begin{equation}\label{eqnorm}
m' = m_{\rm ap} + 2.5\, \log (t_{\rm exp} + \Delta\,t) - k_{\lambda}\,X
\end{equation}
where $m_{\rm ap}$ are the instrumental magnitudes measured in a
circular aperture of radius $R = 6\farcs9$ (SH98), $\Delta\,t$ is the
shutter delay and $X$ is the airmass. For the extinction coefficients
we adopted $k_V = 0.135$ and $k_I = 0.048$ (from the Geneva
Observatory Photometric Group data).

The normalized instrumental magnitudes were then compared to the
Landolt's (\cite{l92}) values, and the following relations were found:

\begin{equation}
V = v' + 0.049 (\pm0.001)\, (V-I) + 23.766
\end{equation}
\begin{equation}
I = i' - 0.007 (\pm0.001)\, (V-I) + 23.070
\end{equation}

where the uncertainties represent the 90\%\ confidence ranges of the
fit for one interesting parameter. The standard deviations of the
residuals are 0.013 mag in V and 0.022 mag in I, respectively.

In order to transform the PSF magnitudes into aperture magnitudes we
assumed that $m_{\rm ap} = m_{\rm PSF} + const.$ (Stetson
\cite{s87}). For each individual frame a sample of bright isolated
objects were then found, and all their neighbors were subtracted. The
`cleaned' images were used to measure aperture magnitudes for the
selected stars, and for each star we computed the difference with
respect to the PSF magnitude obtained on the averaged frames. The same
aperture used for the standard stars was employed. The internal
uncertainty of the calibration of the order of 0.01 mag for each
filter (cf. SH98).

\begin{figure} 
\psfig{figure=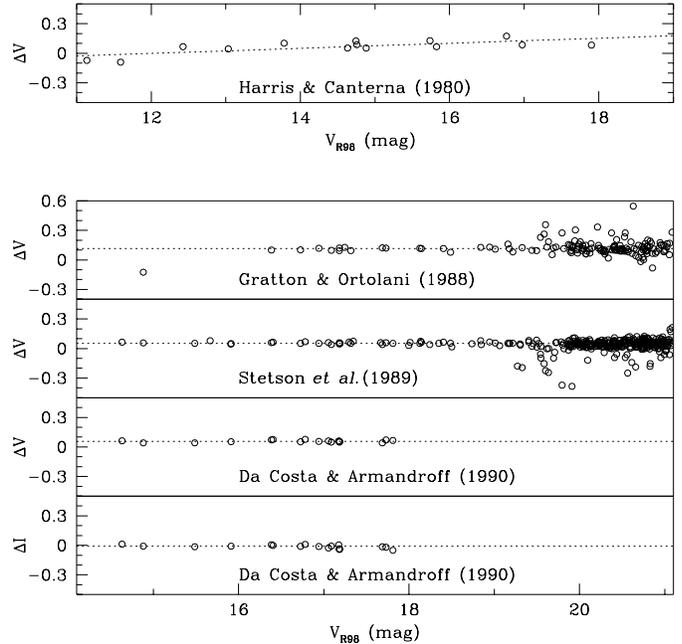,width=8.8cm}
\caption[]{
Comparison with previous photometries. {\it Upper panel.}  Differences
in $V$ between our data and HC80. {\it Lower panels.}  Comparison with
more recent CCD data. The mean difference between our data, GO88 and
S89 is found to be $\Delta V= 0.12\pm0.01$~mag and $\Delta V=
0.05\pm0.01$~mag, respectively. The mean magnitude differences between
our data and DA91 are $\Delta V=0.06\pm0.01$~mag and $\Delta
I=-0.01\pm0.01$.}
\label{comparison}
\end{figure}

Our photometric catalogs are compared with those of Harris \& Canterna
1980 (HC80), GO88, S89 and DA90 in Fig.~\ref{comparison}. Noticeable
differences are found for the $V$ band, where a $\Delta
V\simeq0.05$~mag is present between our values and those of HC80, S89
and DA90, in the sense that our magnitudes are fainter, and an even
larger difference is found between our data and GO88 ($\Delta
V\simeq0.12$~mag, cf Fig.~\ref{comparison}). For the $I$ band, only
the DA90 data allow a comparison, and we find that the two
calibrations match within the errors.

\begin{figure*} 
\psfig{figure=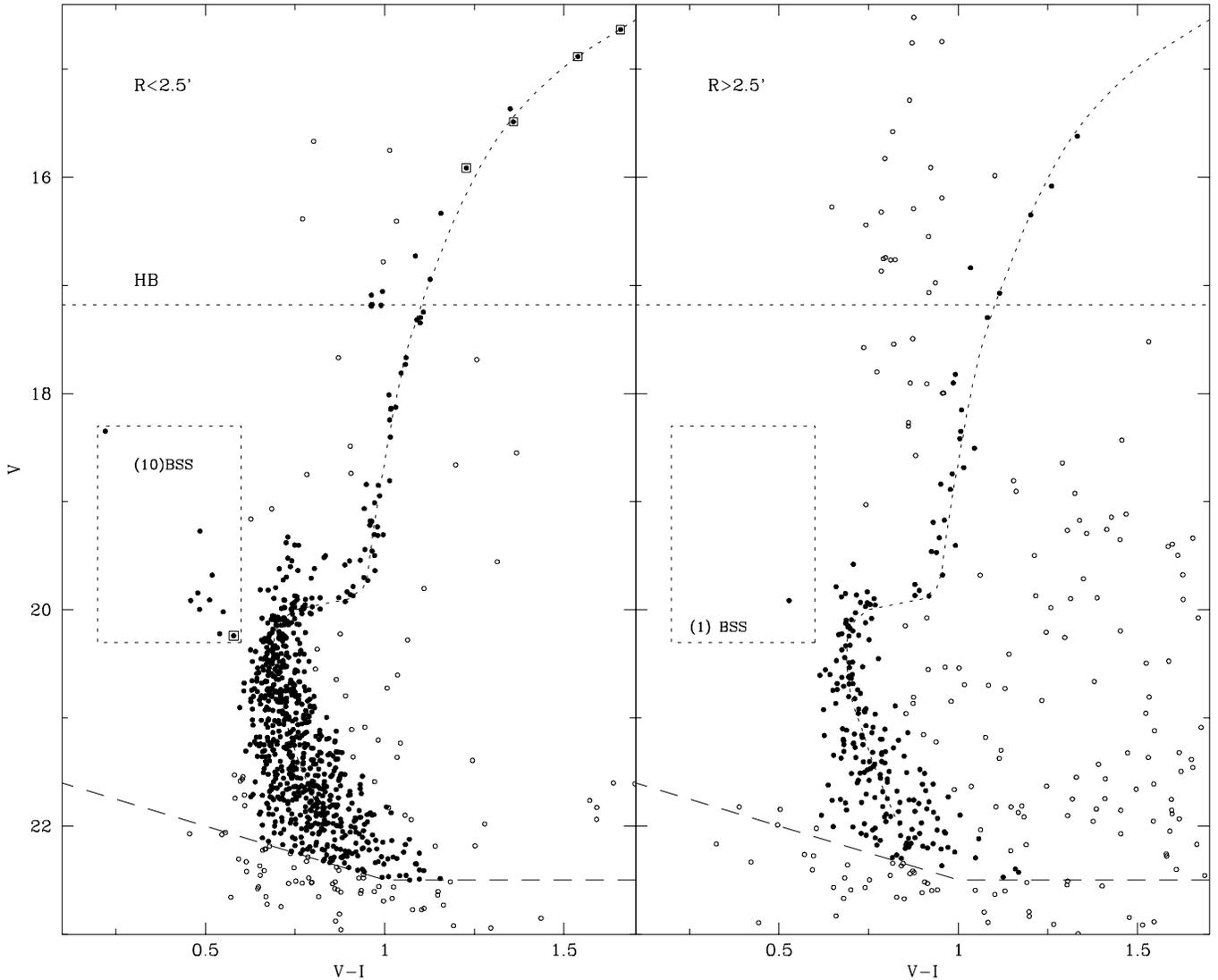,width=18cm}
\caption[]{Color-magnitude diagram for the inner ($R<2\farcm5$, left panel),
and the outer region (right panel) of Palomar 12. The adopted fiducial
points are shown, together with the BSS region. The HB level is
identified by a horizontal dotted line. The stars used for the
computation of the cluster's profile are marked by filled circles. The
$50\%$ completeness level is represented by the dashed line. The four
bright stars in the diagram, marked by squares, are those for which
spectroscopy has been done (DA91, B97). }
\label{cmds}
\end{figure*}

We tried to sort out the possible reason for the observed
discrepancies in the $V$ band, while no significant differences are
found for the $I$ band. From a comparison with existing photometry of
the Fornax dwarf galaxy, SH98 conclude that their $V$ band calibration
is consistent with the previous works. A possible source of
uncertainty could be a problem with the $V$ exposure times: however,
the (small) shutter delay has been included in Equation~1. Moreover,
when the individual zero points of the 7 available $V$ frames are
compared, no differences larger than 0.01~mag are found, which
furtherly confirms that there is no shutter delay problem. In
principle, thin cirrus could have been present at the beginning of the
night, although it should have blocked a remarkably constant
percentage of light during the $\sim 40$~min time span of the cluster
observations, which seems unlikely. The above arguments lead us to
trust our calibration, although further checks are needed in order to
settle this issue.

In any case, the global zero-point difference in $V$ between our
Pal~12 photometry and that of the previous works will not affect our
conclusions on the relative age of this cluster.
\\

\section{The Color-Magnitude Diagram}
\label{sec_cmd}
Fig.~\ref{cmds} shows the CMDs for the stars located inside ({\it left
panel}) and outside ({\it right panel}) the known tidal radius
($r_{\rm t}\simeq2\farcm 5$, Trager et al. \cite{t95}). The main
features of the CMD can be clearly identified also in the outer
region, implying either that the tidal radius must be larger than
previous estimates (cf. Sect.~\ref{sec_struct} for a detailed
discussion), or that Pal~12 is surrounded by a remarkable halo of
extra-tidal radius cluster stars (Grillmair et al. \cite{g95}, Zaggia
et al. \cite{z97}).

\begin{figure} 
\psfig{figure=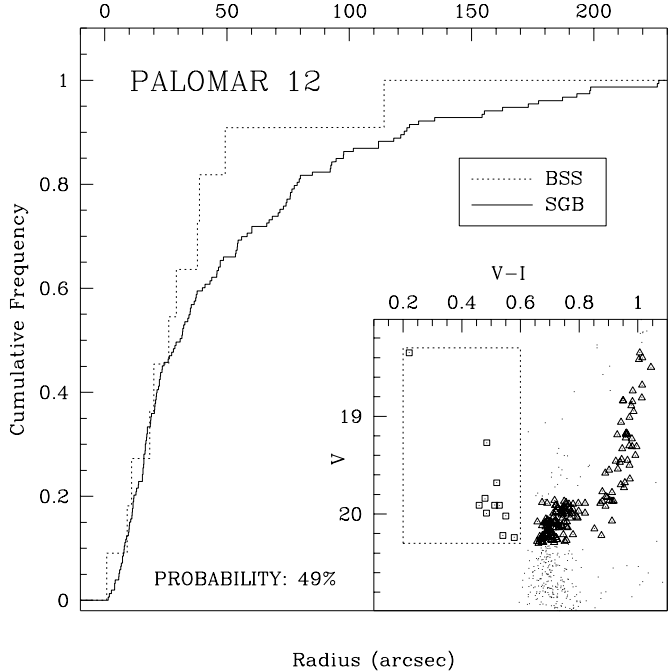,width=8.8cm}
\caption[]{
Cumulative distribution of the BSS and the SGB stars with similar $V$
magnitude in Pal~12. Though the BSS seem to be more concentrated than
the SGB stars, their small number does not allow to assess the
significance of this result. There is a probability of 49\% for the
null hypothesis that the two samples share the same radial
distribution.}
\label{bss}
\end{figure}

Stars from $\sim 2$~mag below the turnoff (TO) up to the red giant
branch (RGB) tip have been measured. Eleven blue straggler stars (BSS)
are clearly identified in the region $0.2<V-I<0.6$,
$18.2<V<20.2$. Nine of them were already known, while two BSS are
located outside the limits of the previously studied fields of the
cluster. The BSS are marked with open triangles in Fig.~\ref{field1}.
As shown in Fig~\ref{bss}, the BSS are more concentrated than the sub
giant branch (SGB) stars with similar magnitude. This is consistent
with what found in other GGCs, though the small number of BSS does not
allow to assess the statistical significance of this result.

In the region $0.60<V-I<0.85$ and $18<V<19.8$ of the inner CMD a
number of stars are present just above the TO. We have compared these
stars with the corresponding objects in the S89 photometry. From this
analysis we found that 45\% of our objects are blends of $2$ S89
stars, 20\% are blends of $3$ S89 stars and 35\% of them are single
stars in the S89 photometry (where the pixel size is just $0\farcs
22$, i.e. half of ours). Notice that almost no such stars are present
in the outer, less crowded, region. It is likely that all stars with
$V<20$ in Fig.~\ref{comparison} being significantly brighter in our
photometry are photometric blends.

The horizontal branch (HB) is formed by 5 stars (already identified in
the literature), and it is located in a very small region around the
point $(V-I,V)= (0.97, 17.18)$, on the red side of the instability
strip, as expected on the basis of the cluster metallicity. A dashed
horizontal line marks the level of the HB in Fig.~\ref{cmds}. The TO
can be identified at $(V-I)=0.69\pm0.01$ and $V=20.50\pm0.1$.

The foreground/background star contamination is low, as expected from
the high galactic latitude of the cluster ($b\simeq-48^\circ$). This
is clearly seen by comparing the right and left panels of
Fig.~\ref{cmds}; the right panel shows the typical pattern of the halo
background, superposed to the cluster CMD. The field contamination is
redder than the Pal~12 MS, and decreases from fainter to brighter
magnitudes. Notice also that the central CMD area is 5 times smaller
that the external one, so that the cluster/background ratio clearly
favors Pal~12 stars.

In order to determine the cluster profile, we defined a sample of
stars with higher membership probability, by selecting all the objects
within $3\sigma$ from the MS-SGB-RGB line (where $\sigma$ represents
the mean error in color as a function of magnitude, as calculated from
the artificial star experiments, and the fiducial line has been drawn
by hand). BSS, HB and photometric blends (see previous discussion)
were added to this sample.

Artificial star tests have been performed in order to investigate the
completeness of our sample. A total of $\sim 60,000$ stars have been
added in $40$ separate runs. The results of these experiments show
that the $50\%$ completeness level is located at $V\simeq22.5$ and
$I\simeq21.5$. Only the stars above these limits (marked by a dashed
line in Fig.~\ref{cmds}) have been selected for the following
analysis. In order to get a meaningful profile it is also critical
that no radial dependence of the completeness exists. We checked that
the completeness profile is constant in the range $R \geq 20$~arcsec
from the cluster center, while a slight rise in magnitude of the
$50\%$ level is observed in the inner region.

The star subsample defined with the previous criteria is identified by
filled circles in Fig.~\ref{cmds}, whereas open circles mark probable
halo field stars. The same convention is used in the maps presented in
Fig.~\ref{field1}.

In order to compare the Pal~12 CMD with other clusters and theoretical
isochrones, a discussion of its relevant parameters is now given.
\\

\subsection{Metallicity}
\label{metallicity}

A summary of early studies on Pal~12 metallicity is presented in
S89. Although a large uncertainty in the metal content determinations
for Pal~12 existed at the time, a combination of several metallicity
indices yielded a value comprised between the ones of M5 and 47~Tuc
(i.e. ${\rm [Fe/H]}=-1.0\pm0.3$).

Since then, three new metallicity determinations have been obtained:
besides new CCD photometry, low and high resolution spectra have been
analyzed for a few giant stars. These stars are marked with open
squares both in the cluster's map (Fig.~\ref{field1}, right panel) and
in the CMD (Fig.~\ref{cmds}, left panel).
 
Da Costa \& Armandroff (\cite{d90}) derived ${\rm
[Fe/H]}=-1.06\pm0.12$ from $V$, $I$ photometry of 20 Pal 12 giant
branch stars, by comparing the position of the RGB with other
calibration clusters. Applying the same method to our data, we obtain
a value ${\rm [Fe/H]} \simeq -0.93$, where the small difference, well
within the uncertainties, is mainly due to our 0.06~mag redder colors
(see Sect.~\ref{obs_reduct}).

Armandroff \& Da Costa (\cite{a91}, DA91) obtained the metallicity
from the Ca~II triplet strenghs, and found ${\rm [Fe/H]} =
-0.60\pm0.14$ for Pal~12, later confirmed by Da Costa \& Armandroff
(\cite{d95}; ${\rm [Fe/H]} = -0.64\pm0.09$).

The most recent result has been obtained by Brown et al. (\cite{b97}).
They present high-resolution spectra of the two brightest stars of
AD91, obtaining a ${\rm [Fe/H]}=-1.0\pm0.1$. They also analyzed the
${\rm [\alpha/Fe]}$ abundances, obtaining a zero value.

In view of the larger uncertainties related to indirect metallicity
determinations with respect to high resolution spectroscopy, in the
following we will adopt ${\rm [Fe/H]}=-1.0$ for Pal~12, and assume a
null $\alpha$ element enhancement.
\\

\subsection{Reddening}
\label{reddening} 

The interstellar reddening towards Pal~12 is expected to be low, given
the high galactic latitude of the cluster. Although no accurate
estimates exist, two independent values have been suggested; HC80
adopted a value of $E(B-V)=0.02\pm0.02$ from the cosecant law (Harris
\& Racine \cite{h79}), and noted that this value is consistent with that
estimated from the color-color diagram of stars in their photoelectric
sequence, $E(B-V)<0.03$. A small reddening is also indicated by the
maps by Burstein and Heiles (\cite{b82}): $0.00<E(B-V)<0.03$. Adopting
$E(V-I)=1.28\,E(B-V)$ (Dean et al. \cite{d78}), we obtain the value
$E(V-I)=0.03\pm0.02$, which will be the assumed reddening throughout
this paper.
\\

\subsection{Distance}
\label{distance}

Distance moduli of the Palomar class clusters have been often
overestimated in the past. Kinman \& Rosino (\cite{k62}) searched
Palomar 12 for variables. They found three variables, one of them
previously discovered by Zwicky (\cite{z57}). Based on the mean
apparent magnitude of these RR Lyrae, Pal 12 was initially located
farther than 50 kpc from the Galactic center (Harris \cite{h76}). It
is only after HC80 photometric study that a more precise distance
modulus has been given (about 14 kpc), on the basis of the $V$
magnitude of the poorly populated HB.

We derive the distance to Pal~12 by comparing its HB with that of
NGC~6362, which is the only GC at ${\rm [Fe/H]}\simeq -1$ with
measured $\alpha$-elements abundance (cf. Tab.~2 in Carney, 1996).
Piotto et al. (\cite{p98}) give $M_V({\rm HB})=\mvhba \pm \errmva$ for
NGC~6362; this value is {\em not} representative of the Pal~12 HB
luminosity, since we must correct for the age (cf. Sect.~\ref{age})
and $\alpha$ abundance offsets between both clusters.

A decrease in age implies an increase in the HB stars mass and
luminosity, the exact dependency being a function of $Z$. Although no
$Z=0.002$ (the Pal~12 metallicity) models are available, an
interpolation from the $Z=0.001$ and $Z=0.004$, Bertelli et
al. (\cite{b94}, hereafter B94) model isochrones leads to estimate a
change $\Delta M_V \sim -0.07$ ~mag, which reduces the age by $30\%$
(cf. Sect.~\ref{age}).

Spectroscopy of 2 NGC~6362 giants has been obtained by Gratton
(\cite{g87}), who measured ${\rm [\alpha/Fe]}=0.32\pm0.09$. In view of
the results by Brown et al. (\cite{b97}) presented in
Sec.~\ref{metallicity}, a comparison of the Pal~12 CMD with NGC 6362
must take into account the ``$\alpha$-enhancement'' of the latter.

As discussed in more detail in Sect.~\ref{age}, an increase of 0.3~dex
in ${\rm [\alpha/Fe]}$ mimics an increase of 0.2~dex in the equivalent
[Fe/H], and implies a decrease in the HB brightness. The exact value
depends on the slope of the luminosity-metallicity relation for the
HB. Although this is still controversial, a typical value $\Delta
M_V/\Delta {\rm [Fe/H]}=0.20$ can be used (Carney et al. \cite{c92}),
which therefore means $\Delta M_V=0.04$~mag in our case.
 
We should also take into account possible differences in the mass loss
rates along the RGB between the two clusters. These would affect the
ZAHB mass, and hence its luminosity. In order to constrain such an
effect, we can compare the colors of the red HB of Pal~12 and
NGC~6362.  Indeed, using again the B94 isochrones we find that, in the
red HB region, a change in the ZAHB mass of +0.1$M_\odot$ will change
the HB location of a star by $+0.22$~mag in $(B-V)$ and $-0.07$~mag in
$V$. The effect is therefore three times larger in the $(B-V)$ color
than in the $V$ magnitude.

The actual dereddened colors of the red HBs of the two clusters are $(B-V)_0
\sim 0.73$ for Pal~12 (Stetson et al. 1989), and $(B-V)_0 \sim 0.54$
for NGC~6362 (Piotto et al. 1998). Hence, a color difference of $\sim
0.2$ mag in $(B-V)$ exists between Pal~12 and NGC~6362, which
corresponds to a $<0.1 M_\odot$ mass loss difference.

However, this higher HB mass for Pal~12 is consistent with its lower
age.  According to B94, the turnoff mass of a cluster will change by
$\sim 0.1 M_\odot$ if its age is changed by $\sim 5$~Gyr. Since, in
the B94 scale, the typical GC age would be $t\sim 14\div15$~Gyr
(Saviane et al. \cite{savetal98}), the higher mass of the Pal~12 HB is
easily explained by its $\sim 30\%$ lower age (cf. Sect.~\ref{age}). A
mass loss differential correction is therefore not needed.

In summary, we expect that the Pal~12 HB should be 0.07~mag brighter
than that of NGC~6362 in view of its younger age and 0.04~mag brighter
due to its lower $\alpha$ element content, i.e. $M_V({\rm HB})=
\mvhbb \pm \errmvb$.

As the apparent magnitude of the Pal~12 HB is $V_{\rm HB} = \vhb \pm
\errvhb$ (where the error has been computed taking into account the
calibration uncertainties), the apparent distance modulus becomes
$m-M_V=\apparent$. Given the assumed reddening $E(B-V)=0.02$, the
absolute distance modulus is $(m-M)_0=\mmzero \pm \errmm$. The
estimate of the error includes the uncertainties on the calibration
zero-point, on the magnitude of the NGC~6362 HB, and on the
absorption. Our value of the distance to Pal~12 is perfectly
compatible with previous estimates: HC80 give $16.2\pm0.35$, GO88
$16.1\div16.5$, S89 $16.3$, and DA90 $16.46$ for the absolute distance
modulus.

The adopted distance modulus corresponds to a distance from the Sun
$R_\odot=\ds\pm\eds$ Kpc, a distance from the Galactic center
$R_{GC}=\dgc\pm\edgc$ kpc, and a height $Z_{GP}=\dgp\pm\edgp$ below
the Galactic plane (we adopted a distance from the Sun to the Galactic
center $R=8.0\pm0.5$ kpc; Reid \cite{r93}).

\begin{figure} 
\psfig{figure=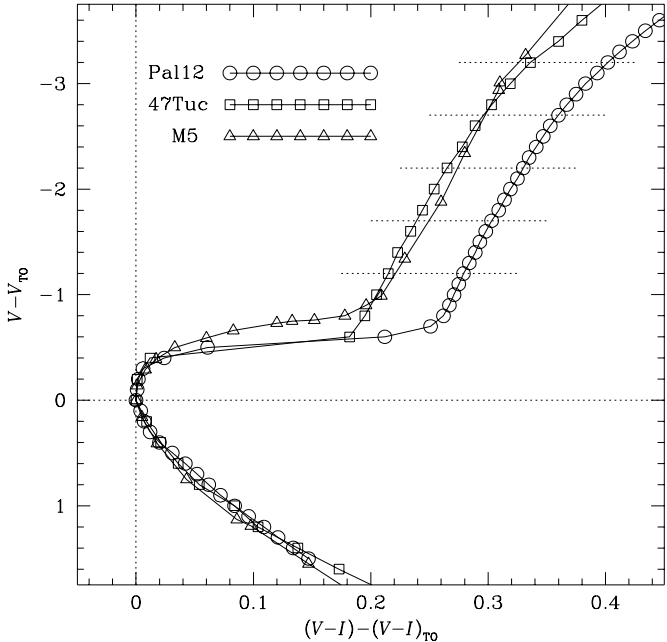,width=8.8cm}
\caption[]{The fiducial points for M5, 47~Tuc and Pal~12 are
presented, after that the TOs have been shifted in magnitude and
colors to a common value. A large difference in color exists between
the RGBs of the two template clusters and the Pal~12 RGB. The small
color difference between the RGB of M5 and the one of 47~Tuc (whose
metallicities encompass that of Pal 12) demonstrates that a 0.5~dex
difference in metallicity has small influence on the $V-I$
color--difference between the RGB and the TO. Pal~12 must therefore be
younger than the two template clusters.}
\label{age1}
\end{figure}

\section{Age}
\label{age}

The first evidence of a relatively young age for Pal~12 was given by
GO88, who estimated that Pal 12 must be 30\% younger than 47~Tuc on
the basis of an atypically small value of the magnitude difference
between the HB and the TO. Indeed, this was the first clear
identification of a young GGC.

Almost at the same time, S89 presented an independent $BV$ CCD
study. They compared the Pal~12 fiducial RGB to those of 47~Tuc and
M5, which bracket Pal~12 metallicity, concluding that no match could
be found. The simplest explanation was that Pal~12 is younger than the
other two clusters by some 25\%-30\%.

\begin{figure} 
\psfig{figure=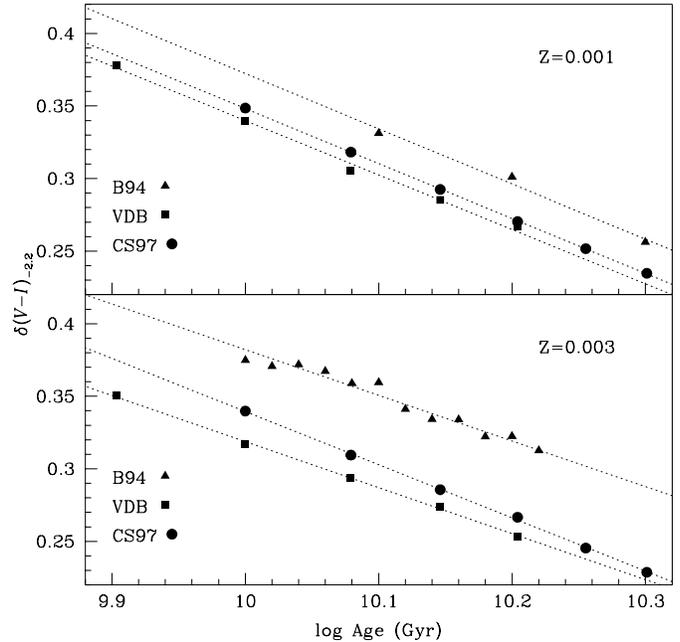,width=8.8cm}
\caption[]{RGB--TO ($V-I$) color-differences at 2.2~mag above the TO 
have been computed using B94, V98, and S97 isochrones, for
different ages and for the two labeled metallicities. Keeping the
metallicity fixed, a linear relation with the same slope (within $\pm
1\%$) is found between the RGB-TO color width and the logarithm of
age, regardless of the model used. The discrepant zero-points are
partly due to different assumptions on the $\alpha$-element content of
the three theoretical sets. The similarity in the slope shows that the
three sets of models give consistent relative ages, at least in the
small metallicity interval considered here.}
\label{age2}
\end{figure}

In both studies, the 47~Tuc fiducial lines were taken from Hesser et
al. (\cite{h87}, hereafter H87). These fiducials were constructed by
merging $B$ and $V$ CCD photometry for 8800 stars below the MS turnoff
to the evolved part of the CMD coming from earlier photographic work
(Hesser \& Hatwick \cite{h77}, Lee \cite{l77}). Also the HB and MS
fiducial lines of M5 come from two different studies ({\it cf.} S89
for more details). Possible photometric calibration discrepancies
between the different datasets contribute to the age uncertainty in
these early estimates.

The heterogeneity of the data base for the comparison clusters and the
high uncertainty in the metal content do not allow us to quantify the
error associated to the results by GO88 and S89. In the following we
will attempt a new, independent determination of the Pal~12 {\em
relative} age by comparison with suitable template clusters.

Since no GGCs with metallicity ${\rm [Fe/H]} \sim -1.0$ have been
observed to date in the $V$, $I$ bands, as done by the previous
authors, we will use 47~Tuc (NGC~104) and M5 (NGC~5904), whose
metallicities bracket that of Pal~12. These are the nearest
metallicity clusters for which (a) published homogeneous $V$, $I$
photometry exists, from the RGB tip down to the MS; (b) both ${\rm
[Fe/H]}$ and $\alpha$-elements abundance have been obtained from
high-resolution spectroscopy; (c) do not show any age anomaly either
in published or in our preliminary analysis of GGC relative ages.

The best $VI$, photometric sample for 47~Tuc is that of Kaluzny et
al. (\cite{k98}, see Paper I for a discussion of the other available
$VI$ CMDs for 47~Tuc). Two photometric catalogs can be used for M5,
namely Sandquist et al. (\cite{s96}) and Johnson \& Bolte
(\cite{j98}). Since Johnson \& Bolte discuss possible problems in
their earlier calibrations of the M5 photometry, we will use the most
recent sample. In any case, the stellar $V-I$ colors are the same in
the two studies.

The metallicities and $\alpha$ abundance ratios have been taken from
Tab.~2 of Carney (\cite{c96}): ${\rm [Fe/H]} = -0.73$ and ${\rm
[\alpha/Fe]}=0.18\pm0.03$ for 47~Tuc, and ${\rm [Fe/H]} = -1.22$ and
${\rm [\alpha/Fe]}=0.30\pm0.03$ for M5.

Figure~\ref{age1} shows the fiducial points of M5, 47~Tuc and Pal~12
registered to a common TO point. It is clear that, while the RGBs of
47~Tuc and M5 are almost overlapping, the RGB of Pal~12 is
significantly redder. The modest color shift between the RGB of M5 and
that of 47~Tuc shows that metallicity differences have small influence
on the RGB--TO color-difference, in the $V$ vs. $V-I$ plane. A change
of 0.5~dex in metallicity implies a color offset as small as
0.01~mag. This fact is confirmed by the theoretical models, and has
been pointed out by Saviane et al. (\cite{s97}). Assuming an age of
14~Gyr, the models of Vandenberg 1998 (hereafter V98) predict a change
of 0.011~mag in $\delta (V-I)$ increasing the metallicity from
$Z=0.002$ to $Z=0.003$ (the color difference between the RGB and the
TO has been measured at 2.2~mag above the TO).

The position of the Pal~12 RGB cannot therefore be explained by a
simple metallicity effect. The observed difference in the location of
the RGB of Pal~12 with respect to 47~Tuc and M5 must be due either to
a different $\alpha$ element abundance or to an age effect.

We begin by examining the first possibility. According to Salaris et
al. (\cite{s93}), an enhancement by a factor $f$ in the ratio
$X_\alpha/X_{\rm Fe}$ is equivalent to an increase of a factor
$(0.638\,f + 0.362)$ in the metallicity $Z$. As discussed in
Sec.~\ref{metallicity}, the current measurements give ${\rm
[\alpha/Fe]}=0$, $0.2$ and $0.3$ for Pal~12, 47~Tuc and M5,
respectively. This means that, in order to compare the Pal~12
fiducials with the reference clusters, we must take into account these
differences in $\alpha$ element abundances, which correspond to
increasing the Pal~12 metallicity by $\sim0.2$~dex. The
$\alpha$-enhancement effect makes the [m/H] of Pal~12 close to that of
47~Tuc. Therefore, Fig.~\ref{age1} shows that the $\alpha$ element
abundance differences cannot justify the large observed RGB color
differences.

An age difference is the only remaining explanation. In order to make
an estimate of the Pal~12 relative age, we have measured $\delta
(V-I)$ between the TO and the RGB for different (fixed) $V-V_{\rm TO}$
values in the models of B94, Straniero et al. (\cite{sc97}, hereafter
S97), and Vandenberg (\cite{v98}, hereafter V98). The first two sets
of models are $non-\alpha-enhanced$, while the third one is.
Figure~\ref{age2} displays the $\delta (V-I)$ for $V-V_{\rm
TO}=-2.2$~mag as a function of the logarithm of age. With a good
approximation, $\delta (V-I)$ linearly depends on the logarithm of
age. The $-2.2$~mag level has been chosen after an analysis of the
behavior of the TO-RGB color difference with respect to the age. We
have repeated our measurements at the RGB levels marked by dotted
lines in Fig.~\ref{age1} and found that, if a value $V-V_{\rm
TO}>-1.2$ is taken, the SGB plays an important role, making relative
measurements difficult to interpret. The same occurs for $V-V_{\rm
TO}<-3.5$, where the slope of the RGB becomes very sensitive to the
clusters metallicity. Conversely, for $V-V_{\rm TO}$ in the range
$[-1.2\div-3.2]$ and age older than 8 Gyrs, the $\delta (V-I)$ seems
to be almost independent of metallicity. We simply chose a mean value
$-2.2$. The linear relations in Fig.~\ref{age1} have the same slopes
for $Z=0.001$, while for $Z=0.003$ the B94 and V98 models give the
same slope, which is slightly different from that obtained from
S97. The zero points are different, but this does not affect the
relative age determination. We will therefore obtain the same relative
ages when using either the B94, V98 or S97 models at $Z=0.001$, while
the S97 isochrones give age differences larger by $\sim 4\%$ than B94
or V98 at Z=0.003.

From Fig.~\ref{age1} we have $\delta (V-I)$=0.280 for M5, $\delta
(V-I)$=0.265 for 47~Tuc, and $\delta (V-I)$=0.330 for Pal 12.
Assuming Z=0.003, from Fig.~\ref{age2}, we obtain that Pal 12 is 34\%,
34\%, or 30\% younger than 47 Tuc on the basis of V98, B94 and S97
models, respectively. As discussed above, adopting Z=0.001 we have
quite similar results: formally, Pal 12 is 33\%, 32\%, or 32\% younger
than M5. Taking into account the errors in measuring the $\delta
(V-I)$ parameter (estimated assuming an uncertainty of $\pm0.15$ mag
and $\pm0.01$ mag in the magnitude and color of the TO) for both Pal
12 and the reference clusters, the uncertainties in the relative ages
is of the order of 10\%. We conclude that Pal~12 has an age $68\% \pm
10\%$ that of a typical GGC, assuming that 47~Tuc and M5 age are
representative of the ages of the bulk of the GGC population (Buonanno
et al \cite{b98}).
 
\section{Structural Parameters}
\label{sec_struct}

\begin{figure} 
\psfig{figure=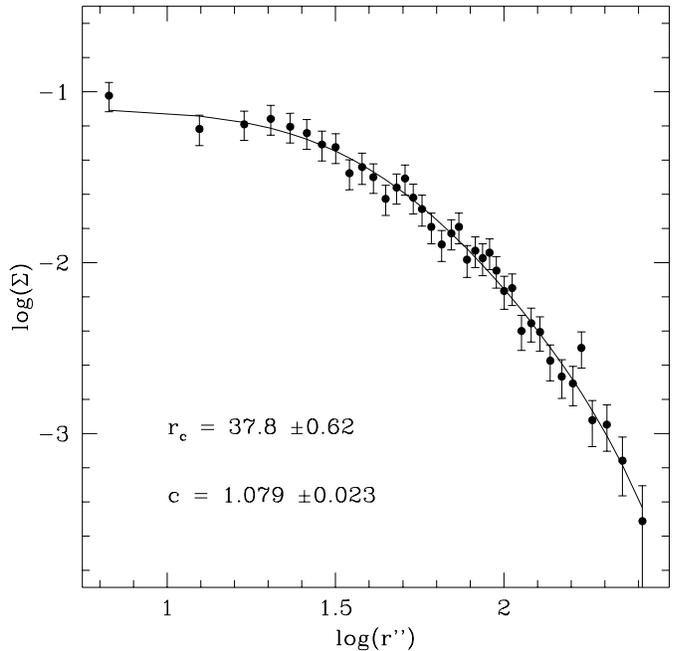,width=8.8cm}
\caption[]{Profile determination for Palomar 12; filled
circles represent the observed star counts and the solid line the
best-fitting King law. The adopted morphological parameters are
$r_{\rm c} = \rc \pm \errrc$ and $c = \conc \pm \errc$. 
}
\label{profilo}
\end{figure}

We have derived the density profile of Pal~12 making radial star
counts in equal number-of-stars steps. Stars with high membership
probability were selected as explained in Sect.~\ref{sec_cmd}. The
cluster center was determined in an iterative manner, by first
computing a median of the $x$ and $y$ coordinates of the stars within
an arbitrarily located circle of radius $r=160\arcsec$. Next, a new
circle was considered with its center corresponding to the just
obtained median point. The process was repeated until two subsequently
computed centers were coincident. The offset between our center and
that given by HC80 is just $\dx \arcsec$ in RA and $\dy \arcsec$ in
DEC. The background level was estimated outside a suitably large
distance from the cluster center, chosen with the following procedure.
First, a King law was fitted to the observed profile, and a set of
structural parameters was derived. Then the procedure was repeated
until the computed tidal radius was smaller than the radius used for
the foreground estimate.

The final result is shown in Fig.~\ref{profilo}, where the filled
circles represent the observed star counts and the solid line the
best-fitting King law. The adopted morphological parameters are
$r_{\rm c} = \rc \pm \errrc$ and $c = \conc \pm \errc$, where the
errors were estimated following the methods adopted by Saviane et
al. (\cite{s__96}), and represent the formal uncertainties of the fit.

A more reliable estimate of the errors was computed by keeping the
central density fixed and varying the other 2 parameters in a grid of
values. The 90\%\ confidence ranges of the fit for the tidal radius
and concentration are $r_{\rm c} = \rc^{\rcup}_{\rclow}$ and $c =
\conc^{\concup}_{\conclow}$. An alternative estimate of the
uncertainties of the parameters was obtained by changing the sky level
by $\pm3\sigma$. The effect is to lower $r_{\rm c}$ by $\sim
1.5\arcsec$ and to change $c$ by $\sim \pm 0.1$.

Our structural parameters are significantly different from those
published in Trager et al. (\cite{t95}); the authors quote $\log
r_{\rm c}=0 \farcs 23$ and $c=1.94$, which imply a tidal radius
$r_{\rm t}=2\farcm 46$. Since stars belonging to Pal~12 are clearly
seen beyond this radial limit (cf. Fig.~\ref{cmds}), it is clear that
the Trager's et al. tidal radius is too small. On the other hand, the
same authors had previously listed values closer to ours (Trager et
al. \cite{t93}), analyzing the same data used by Trager et
al. (\cite{t95}). It is therefore possible that the unconsistency
comes from some typo in Trager et al. (\cite{t95}) table.

As a final remark, we notice that a dip in the profile is observed in
the very central region. This could be due to a slightly lower
completeness as discussed in Sect.~\ref{sec_cmd}. We fitted the
profile also removing the central 3 points, obtaining almost identical
structural parameters.

\section{Summary and conclusions}
\label{conclusions}

The first deep $V$, $I$ CCD photometry for the Galactic globular
cluster Palomar~12 has been presented. The wide field allowed us to
sample the cluster stellar population well beyond the tidal radius.
All stars in our field down to $\sim 2$~mag below the MS turnoff
($50\%$ completeness level) have been measured, allowing a clear
definition of all the CMD sequences.

Using the HB brightness, an improved distance determination has been
obtained by comparison with NGC~6362 as a reference cluster.

A direct comparison with homogeneous $V$, $I$ CMDs for 47~Tuc and M5
shows that Pal~12 is a young cluster. The computation of a precise
relative age depends on the theoretical isochrones used, although
differences of at most $4\%$ are found among the three models
considered (B94, V98 and S97). The comparison with the models shows
that Pal~12 age is $68\% \pm 10\%$ that of the reference clusters.

Finally, our large field also allowed to obtain a radially complete
number density profile for stars brighter than $V=22.5$, and to
compute improved structural parameters. The new morphological
parameters are $r_{\rm c} = \rc \pm \errrc$ and $c = \conc \pm \errc$.

\end{document}